\documentclass[prl,twocolumn,superscriptaddress,showpacs,floatfix]{revtex4-2}

\usepackage[utf8]{inputenc}
\usepackage{graphicx}
\usepackage{dcolumn}
\usepackage{braket}
\usepackage{bm}
\usepackage{amsfonts} 
\usepackage{amsmath} 
\usepackage[bookmarksnumbered,pdfpagelabels=true,plainpages=false,colorlinks=true,linkcolor=blue,citecolor=blue,urlcolor=blue]{hyperref}
\usepackage[bookmarksnumbered,pdfpagelabels=true,plainpages=false,colorlinks=true,linkcolor=blue,citecolor=blue,urlcolor=blue]{hyperref}

\def \usach {Departamento de F\'isica, CEDENNA, Universidad de Santiago de Chile, 9170124, Santiago, Chile.}

\begin{document}

 \title{Vacuum-selected timescales in driven Josephson systems.}

\author{Sebastian Allende}
\affiliation{\usach}

\author{David Galvez-Poblete}
\affiliation{\usach}

\begin{abstract}
In this work, we demonstrate that the intrinsic time scale of a Josephson junction can be controlled through dynamical vacuum selection. By applying a Kapitza-like high-frequency drive to the system, the effective Josephson potential is reshaped, allowing for the stabilization of inphase or antiphase configuration. As a result, the Josephson plasma frequency, that is, the clock frequency of the junction, becomes a tunable property of the selected vacuum. Our findings establish a vacuum-controlled Josephson clock principle, in which the dynamical vacuum acts as an internal reference that fixes the operational timescale of Josephson oscillations, rather than this scale being imposed externally.
\end{abstract}

\maketitle

\section*{Introduction}

In 1962 Brian Josephson predicted that a dissipation-less supercurrent can flow through a weak link via quantum tunneling between two macroscopic reservoirs \cite{Josephson1962}, revealing phase coherence at the macroscopic scale in systems such as superconductors \cite{Anderson1963,Shapiro1963, Ambegaokar1963} and Bose-Einstein condensates (BECs) \cite{Abad2011,Adhikari2014,Giovanazzi2000,Levy2007,Valtolina2015,PhysRevA.59.620}. In the low-energy regime, the relevant dynamical variable is the relative phase between the reservoirs \cite{Likharev1979, Su2015}, and the characteristic timescale of the process is set by the Josephson plasma frequency, which is determined by the tunneling properties of the junction \cite{Golubov2004, Pigneur2018}. The Josephson effect is a cornerstone of modern superconducting circuits \cite{Devoret2013,Clarke2008,Blais2021,Stewart1968,2004}, and achieving precise and flexible control over its dynamics remains a central challenge. From a physical perspective, this also implies that the characteristic time scale of the Josephson dynamics, set by the plasma frequency, is not necessarily an immutable property of the junction, but may depend on the configuration around which the phase dynamics is organized.

In standard Josephson junctions, the low-energy dynamics is typically organized around an in-phase ground-state configuration, $\theta_1-\theta_2 =0$, which in turn sets the reference plasma frequency of the junction. In recent years, Floquet engineering has emerged as a powerful strategy to control dynamical properties by applying high-frequency drives that effectively reshape the potential governing slow degrees of freedom \cite{Bukov2015,Ji2022,Lin2025}, in close analogy with the Kapitza mechanism in classical systems\cite{kapitza1951pendulum}. While several driven platforms have demonstrated substantial tunability of Josephson parameters, a clear route to control the intrinsic timescale of the junction through drive-induced vacuum selection remains less explored.  This raises a more general question: can the intrinsic timescale accessed through Josephson dynamics be controlled by dynamically selecting the vacuum configuration of the system? Here, by vacuum we mean a dynamically stabilized minimum of the effective Josephson potential that organizes the low-energy phase dynamics. In this context, it is important to note that the notion of a Josephson clock only emerges once the phase dynamics organizes around a specific stabilized vacuum.

In this work, we propose a route to control the Josephson plasma frequency, and hence the characteristic timescales of the dynamics, by applying a high-frequency drive that enables vacuum tuning of the junction. In particular, the drive can stabilize an antiphase ground state, realizing an effective $\pi$-Josephson junction. We derive an effective equation of motion in the high-frequency limit, obtain the corresponding driven Josephson potential, and identify the stability conditions under which the $\pi$ vacuum emerges. We further analyze linear excitations about each vacuum and how the plasma frequency, and thus the Josephson "clock rate", becomes a controllable quantity.

\section{Theoretical Model}

We consider a system described by an order parameter $\Psi(\textbf{r},t)$:
\begin{equation}
    \Psi (\textbf{r},t) = \sqrt{n(\textbf{r},t)} \text{e}^{i \theta(\textbf{r},t)}
\end{equation}
This order parameter may represent the Cooper-pair wavefunction in a superconductor \cite{Tinkham2004-iq} or the condensate wavefunction in a Bose-Einstein condensate \cite{Pitaevskii2016}. In the latter case, the system can be described by the Lagrangian density \cite{Pethick2008}:
\begin{equation}
    \mathcal{L}[\Psi(\textbf{r},t)] = i \hbar \Psi^* \partial_t \Psi -\frac{\hbar^2}{2m}|\nabla \Psi |^2- \frac{g}{2}|\Psi|^4 + \mu |\Psi |^2
\end{equation}
where $m$ is the particle mass, $g$ the contact interaction strength, and $\mu$ the chemical potential. Introducing the density-phase representation of the order parameter into the lagrangian density and expanding the density around its background value, $n(\textbf{r},t) = n_o(\textbf{r})+\delta n(\textbf{r},t)$, with $|\delta n| \ll n_o$, we retain terms up to quadratic order in $\delta n$ and integrate out $\delta n$, obtaining an effective low-energy Lagrangian for the phase field $\theta$ \cite{Stoof2009-lj}:

\begin{equation}
    \mathcal{L}_{\text{eff}}[\theta] = \frac{\chi }{2} \dot{\theta}^2 - \frac{\rho}{2}(\nabla \theta ) ^2,
\end{equation}

where $\chi$ is the phase compressibility (inertia) and $\rho$ is the phase stiffness. The effective Lagrangian derived above is valid for smooth variations of the phase field. It describes a system with a gapless (phononic) dispersion relation. We now consider two Bose-Einstein condensates in a standard Josephson-junction configuration. In this case, the same hydrodynamic description applies to the relative phase $ \theta (\text{r},t) \equiv \theta_1(\textbf{r},t) - \theta_2 (\textbf{r},t)$, but the effective Lagrangian acquires an additional contribution originating from single-particle tunneling betweeen the condensates. This contribution follows from the tunneling Hamiltonian and introduces a Josephson coupling term \cite{Nagaosa1999}:
\begin{equation}
    \mathcal{H}_{12} = - t \left(\Psi_1^* \Psi_2 + \text{h.c}  \right) \to -J \cos (\theta)
\end{equation}
With this inclusion, the effective Lagrangian governing the Josephson dynamics of the relative phase reads:
\begin{equation}
    \mathcal{L}_{\text{eff}}[\theta] = \frac{\chi }{2} \dot{\theta}^2 - \frac{\rho}{2}(\nabla \theta ) ^2 + J \text{cos}(\theta )
\end{equation}
The equation of motion for this system is a Sine-Gordon equation \cite{Altland2010}:
\begin{equation}
    \ddot{\theta}- \frac{\rho}{\chi } \nabla^2 \theta + \frac{J}{\chi} \sin(\theta) =0
\end{equation}
And in the linear regime, the corresponding dispersion relation is:
\begin{equation}
    \omega^2(\textbf{k}) = c^2 \textbf{k}^2 + \omega_J^2 ; \quad \omega_J^2 \equiv \frac{J}{\chi}  
\end{equation}

To study the dynamical response of the system under a fast Kapitza-like drive, we introduce a time-periodic modulation of the tunneling coupling (Josephson), $J = J(t) = J_o + J_1 \cos(\Omega t)$, with $\Omega \gg \omega_J$.

\begin{equation}
    \ddot{\theta}- \frac{\rho}{\chi } \nabla^2 \theta + \frac{1}{\chi}\left( J_o + J_1 \cos (\Omega t) \right) \sin(\theta) =0
\end{equation}

This modulation introduces two distinct timescales in the dynamics of $\theta(t)$:
\begin{equation}
    \theta (t) = \phi(t) + \xi(t) ; \quad \xi(t) = \Gamma(t) \cos(\Omega t)
\end{equation}

\begin{equation}
    \sin (\phi(t) + \xi(t)) \approx \sin (\phi(t)) + \xi(t) \cos(\phi(t))
\end{equation}

Here $\phi(t)$ denotes the slow component of the phase dynamics, with $\dot{\phi} \sim \omega_J$, whereas $\xi(t)$ is a small rapidly oscillating correction satisfying $\dot{\xi}(t) \sim \Omega$. The rapidly oscillating component $\xi(t)=\Gamma\cos(\Omega t)$ oscillates at frequency $\Omega$, while its amplitude $\Gamma$ is small and scales as $\Gamma=\mathcal{O}(J_1/\chi\Omega^2)$. Consequently, the leading drive-induced correction to the slow dynamics appears at order $\mathcal{O}(J_1^2/\chi\Omega^2)$. For simplicity, we restrict ourselves to spatially uniform phase dynamics. Substituting the decomposition into the equation of motion and performing a harmonic analysis, we find that $\dot{\Gamma}(t) =0$, and:
\begin{equation}
    \Gamma = \frac{J_1}{\chi \Omega^2} \sin(\phi)
\end{equation}

With this, the equation of motion becomes:
\begin{multline}
  \chi  \ddot{\phi}(t) +  
  \left[ \frac{J_1^2}{\chi \Omega^2} \cos^2(\Omega t) + \frac{J_o J_1}{\chi \Omega^2} \cos(\Omega t)\right] \sin(\phi) \cos(\phi) + \\
  J_o \sin (\phi )=0    
\end{multline}

We can average over the fast oscillations:
\begin{equation}
    \langle \cos (\Omega t) \rangle \to 0 ; \quad \langle \cos^2 (\Omega t) \rangle \to 1/2
\end{equation}

With this averaging procedure, we finally obtain an effective equation of motion for the relative phase:
\begin{equation}
    \chi \ddot{\phi} + J_o \sin(\phi) + \frac{J_1^2}{2 \chi \Omega^2} \sin (\phi) \cos(\phi) =0
\end{equation}

where the last term accounts for a drive-induced dynamical correction to the Josephson dynamics. Consequently, the effective Josephson potential can be written as:
\begin{equation}
    V_\text{eff} =  -J_o \cos(\phi) + \frac{J_1^2}{4  \chi \Omega^2} \sin ^2 (\phi) 
\end{equation}

It is woth noting that, in the absence of driving, the previous potential exhibits only two stationary configurations: a minimum at $\phi =0$ and a maximum at $\phi = \pi$. In contrast, once the drive-induced dynamical correction is included, the potential develops three stationary points:

\begin{equation}
    \phi= \{ 0, \pi, \cos^{-1}\left( \frac{-2 J_o \chi \Omega^2 }{J_1 ^2}\right)\}
\end{equation}

Where the third stationary point may or may not exist, depending on the system parameters ($\left|2\chi\Omega^2\, J_0/J_1^2\right|\le 1$), and it always corresponds to a maximum of the effective potential. The configuration $\phi = 0$ remains a minimum, whereas $\phi = \pi$ can become either a maximum or a minimum depending on the driving parameters. To determine its stability, we evaluate the second derivative of the effective potential at $\phi = \pi$:
\begin{equation}
    V_\text{eff} ''  (\phi = \pi ) = -J_o + \frac{J_1^2}{2 \chi \Omega^2}
\end{equation}

Specifically, $\phi = \pi$ corresponds to a minimum if:
\begin{equation}
    \frac{J_1^2}{2 \chi \Omega^2} > J_o
\end{equation}

\begin{figure}
    \centering
    \includegraphics[width=1.0\linewidth]{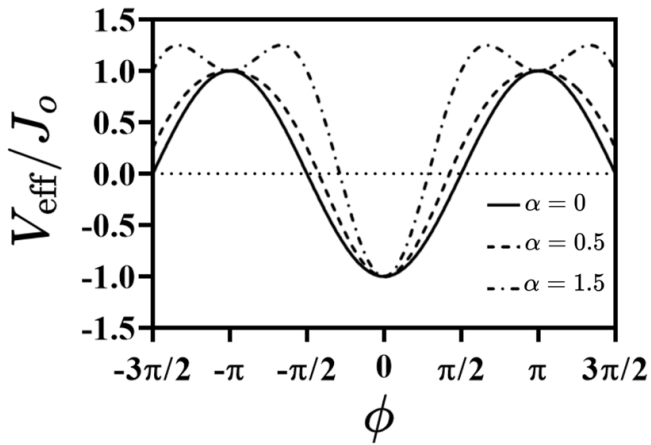}
    \caption{Dimensionless effective potential $V_{\text{eff}}/J_o$ as a function of the phase $\phi$, for different values of the driving parameter $\alpha \equiv J_1^2/2\chi J_o \Omega^2$. The case $\alpha=0$ corresponds to the undriven system ($J_1=0$). For $\alpha=0.5$, the driving is insufficient to stabilize the antiphase configuration. For $\alpha =1.5$, satisfying the  condition of Eq. (18), the potential exhibits two minima at $\phi =0$ (inphase) and $\phi = \pi$ (antiphase).}
    \label{fig1}
\end{figure}

Under this condition, the system's minimum-energy configuration can occur in the antiphase state, $\phi =\pi$, see Fig. \ref{fig1}.

To study fluctuations about the antiphase state, we set $\phi (\textbf{r},t) = \pi + \varphi(\textbf{r},t)$, with $|\varphi| \ll1$, and linearize the equation of motion to first order in $\varphi$, yielding:

\begin{equation}
\chi \ddot{\varphi} - \rho \nabla^2 \varphi + \left( \frac{J_1^2}{2 \chi \Omega^2} - J_o \right)  \varphi =0
\end{equation}

Assuming plane-wave (harmonic) excitations of the form $\varphi (\textbf{r},t ) = A \text{e}^{i(\textbf{k} \cdot \textbf{r} - \omega t)} $, we obtain the dispersion relation:
\begin{equation}
    \omega^2 (\textbf{k}) = \frac{\rho}{\chi}\textbf{k}^2 + \frac{1}{\chi}\left( \frac{J_1^2}{2 \chi \Omega^2} - J_o \right)   
\end{equation}

The corresponding energy density of these excitations associated with the slow component, which we refer to as $\pi$-Goldstone-like modes, is:
\begin{equation}
    \varepsilon[\varphi] = \frac{\chi}{2} (\partial_t \varphi)^2 + \frac{\rho}{2}(\nabla \varphi)^2 - \frac{J_o}{2} \varphi^2
\end{equation}

We consider small real fluctuations described by $\varphi = A \cos(\textbf{k}\cdot \textbf{r}-\omega t )$ and taking the time average over one oscillation period, we obtain:
\begin{equation}
    \langle \varepsilon \rangle = \frac{|A|^2}{4} \left( \chi \omega^2(\textbf{k}) + \rho \textbf{k}^2   - J_o  \right)
\end{equation}

Substituting the dispersion relation $\omega(\textbf{k})$ obtained above, we finally arrive at:
\begin{equation}
        \langle \varepsilon \rangle = \frac{|A|^2}{2} \left( \rho \textbf{k}^2+ \frac{J_1^2}{4\chi \Omega^2}-J_o \right)
\end{equation}

It is worth emphasizing that the energy density of the phase fluctuations is defined relative to the selected dynamical vacuum. In this sense, for a certain parameter regime, $|\mathbf{k}|<k_c$, long-wavelength fluctuations reduce the energy density relative to the selected dynamical vacuum.  In this respect, these excitations are analogous to antimagnons \cite{Harms2024} or antiferron-like modes \cite{GalvezPoblete2025}, in the sense that they reduce the energy relative to a metastable reference vacuum without implying an unbounded spectrum. From the expression above, we can therefore derive the parameter regime in which these excitations lower the energy of the system, together with the stabilization condition required for the antiphase state:


\begin{equation}
    2J_o < \frac{J_1^2}{\chi \Omega^2}<4 J_o
\end{equation}

and the corresponding critical wavevector:
\begin{equation}
    k_c = \sqrt{\frac{1}{\rho}\left( J_o - \frac{J_1^2}{4\chi \Omega^2} \right)}
\end{equation}

\section*{Vacuum-dependent Josephson clock}
Consider a standard Josephson junction in the spatially uniform limit, where $\theta(\textbf{r},t) = \theta(t)$. Under the high-frequency modulation of the tunneling coupling, the effective Lagrangian takes the form:
\begin{equation*}
    \mathcal{L}_{JJ} =  \frac{C}{2} \left( \frac{\hbar}{2e} \right)^2 \dot{\theta}^2 + (J_o +J_1 \cos(\Omega t) )\cos(\theta)
\end{equation*}

With the identification $\chi \equiv C \left( \frac{\hbar}{2e} \right)^2$. As discussed above, this system can exhibit two local minima of the effective potential, the in-phase configuration $\theta =0 $, which is always stable, and the antiphase configuration $\theta = \pi$, whose stability depends on the condition given by Eq. (18). The dispersion relations for small excitations around these configurations are:
\begin{equation}
    \omega^2 (\textbf{k}, \theta =0) = \frac{\rho}{\chi}\textbf{k}^2 + \frac{1}{\chi}\left( \frac{J_1^2}{2 \chi \Omega^2} + J_o \right)   
\end{equation}

\begin{equation}
    \omega^2 (\textbf{k}, \theta =\pi) = \frac{\rho}{\chi}\textbf{k}^2 + \frac{1}{\chi}\left( \frac{J_1^2}{2 \chi \Omega^2} - J_o \right)   
\end{equation}

\begin{figure}
    \centering
    \includegraphics[width=1.0\linewidth]{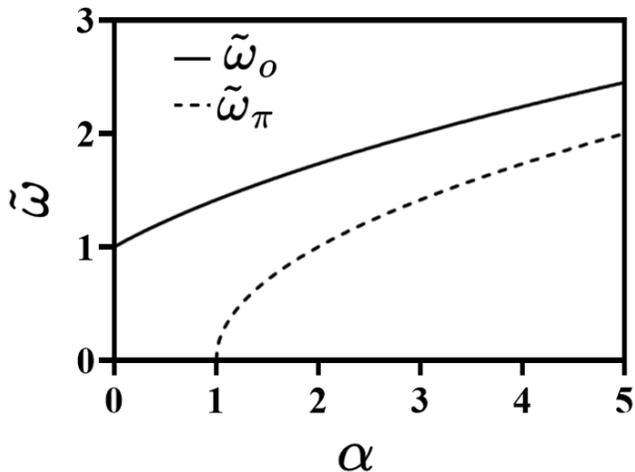}
    \caption{Dimensionless Josephson plasma frequency $\tilde{\omega} = \omega\sqrt{\chi/J_o}$ versus $\alpha =  J_1^2/2\chi J_o \Omega^2$ for the inphase and antiphase dynamical regimes.}
    \label{fig2}
\end{figure}

We define the plasma frequency at $\textbf{k}=0$. Since the system admits two possible vacua (depending on the stability conditions), we obtain two distinct plasma frequencies, corresponding to fluctuations about $\theta =0$ and $\theta = \pi$, respectively: 
\begin{equation}
    \omega_{0 } = \sqrt{\frac{1}{\chi}\left( \frac{J_1^2}{2 \chi \Omega^2} + J_o \right)} ; \quad  \omega_{\pi } = \sqrt{\frac{1}{\chi}\left( \frac{J_1^2}{2 \chi \Omega^2} - J_o \right)}, \quad \label{fesdf}
\end{equation}
see Fig. \ref{fig2}. The plasma frequency sets the intrinsic timescale of the junction dynamics. In this case, the observed effect is not a Floquet renormalization. The frequencies in Eq. \ref{fesdf} are not obtained by renormalizing parameters within a single vacuum, but instead emerge from the dynamical selection of distinct vacua, as reflected in Eq. \ref{fesdf}. Consequently, the characteristic timescale can be tuned by applying a high-frequency drive that controls which minimum is realized. This constitutes the vacuum-controlled “Josephson clock” principle. This effect does not rely on microscopic details of the junction, but follows generically from vacuum selection under high-frequency driving. As a result, two otherwise identical junctions subjected to the same external drive but prepared in different stabilized vacua can exhibit different intrinsic clock rates solely due to their vacuum configuration, accumulating a relative phase offset through their vacuum-dependent plasma frequencies. This effect cannot be interpreted as a mere renormalization of coupling constants, but rather as a manifestation of vacuum-controlled dynamical organization.

More generally, In coherent systems whose slow dynamics is organized around dynamically stabilized vacua, the intrinsic operational timescale is not fixed a priori by external parameters, but is selected by the vacuum configuration around which the dynamics is defined. In Josephson systems, this operational timescale is set by the plasma frequency, which is determined by the curvature of the effective potential at the selected vacuum. Consequently, distinct dynamically stabilized vacua correspond to inequivalent intrinsic clock rates, even under identical external driving conditions.

\section*{Vacuum-selected temporal order toy model}

Building on the vacuum-selected clock principle introduced above, this toy model highlights a route to dynamically control intrinsic time scales in coherent systems, suggesting applications where temporal functionality is encoded in vacuum selection rather than parameter tuning. The central idea of this toy model is to show how a time-crystal–like temporal order can emerge, not because the drive imposes a subharmonic oscillation, but because the system alternates between inequivalent dynamical vacua, each with its own intrinsic timescale. When the system under the drive has the condition $\alpha > 1$, we have:

\begin{equation}
\omega_{\mathrm{clock}} =
\begin{cases}
\omega_0, & \text{Vacuum A} \; (\theta = 0),\\[4pt]
\omega_\pi, & \text{Vacuum B} \; (\theta = \pi).
\end{cases}
\end{equation}

 We already know that the selected vacuum sets the internal clock of the system. Now, by introducing a weak stroboscopic bias, their degeneracy is periodically lifted, thereby alternating which vacuum organizes the slow dynamics, with period $T$. In other words, the vacuum becomes a discrete dynamical variable ($\sigma_n$),i.e.: 
 
 \begin{equation}
\sigma_n =
\begin{cases}
+1, & \text{Vacuum A} \; (\theta = 0),\\[4pt]
-1, & \text{Vacuum B} \; (\theta = \pi).
\end{cases}
\end{equation} 

With the weak stroboscopic bias, the system to be in Vacuum A during odd driving cycles and in Vacuum B during even driving cycles, $\sigma_{n+1} = -\,\sigma_n$. Consequently, the sequence of vacua is “$A$, $B$, $A$, $B$, ...” with a vacuum periodicity of $2T$,i.e., $\sigma(t + 2T) = \sigma(t)$. Therefore, the system breaks the discrete temporal symmetry of the drive in a stroboscopic sense through the identity of the vacuum. Furthermore, each time the vacuum changes, the internal clock of the system changes, causing the internal timescale itself to oscillate,i.e., $\tau_{\mathrm{int}}(t)=\frac{2\pi}{\omega_{\mathrm{clock}}\!\left[\sigma(t)\right]}$.

\section*{Conclusions}

This work demonstrates that the Josephson plasma frequency should be regarded as a vacuum-dependent timescale rather than a fixed parameter of the junction. In a driven Josephson system, distinct dynamically stabilized vacua organize the phase dynamics around different curvatures of the effective potential, leading to inequivalent intrinsic timescales.

With these results, we show that when time is defined operationally through Josephson oscillations, the intrinsic clock rate governing the dynamics is selected by the dynamical vacuum around which the driven Josephson system is organized, even under identical external driving. More generally, this work highlights vacuum selection as a fundamental mechanism for controlling temporal properties in coherent quantum systems, independent of microscopic details. Our results open a route toward a vacuum-indexed form of temporal order, in which temporal organization is encoded in the choice of dynamical vacuum rather than in an explicit breaking of the periodicity of the drive.

\section*{Acknowledgement}

S.A. acknowledges funding from DICYT regular 042431AP and CEDENNA CIA250002. D. G.-P. acknowledges ANID-Subdirección de Capital Humano/Doctorado Nacional/2023-21230818.

\end{document}